\documentclass[
 amsmath,amssymb,
 aps,
 prc,
 twocolumn,
 superscriptaddress,
 mathtools
]{revtex4-2}

\usepackage{graphicx}
\usepackage{dcolumn}
\usepackage{bm}
\usepackage{xcolor}
\usepackage[caption=false]{subfig}
\usepackage{mwe}

\begin{document}

\title{Improved Nuclear Physics Near $A=61$ Refines Urca Neutrino Luminosities in Accreted Neutron Star Crusts}

\author{Z.~Meisel}
\email[]{meisel@ohio.edu}
\affiliation{Institute of Nuclear \& Particle Physics, Department of Physics \& Astronomy, Ohio University, Athens, Ohio 45701, USA}

\author{A.~Hamaker}
\affiliation{National Superconducting Cyclotron Laboratory, East Lansing, Michigan 48824, USA}
\affiliation{Department of Physics \& Astronomy, Michigan State University, East Lansing, Michigan 48824, USA}

\author{G.~Bollen}
\affiliation{Department of Physics \& Astronomy, Michigan State University, East Lansing, Michigan 48824, USA}
\affiliation{Facility for Rare Isotope Beams, East Lansing, Michigan 48824, USA}

\author{B.A.~Brown}
\affiliation{National Superconducting Cyclotron Laboratory, East Lansing, Michigan 48824, USA}
\affiliation{Department of Physics \& Astronomy, Michigan State University, East Lansing, Michigan 48824, USA}
\affiliation{Joint Institute for Nuclear Astrophysics -- Center for the Evolution of the Elements, Michigan State University, East Lansing, Michigan 48824, USA}

\author{M.~Eibach}
\affiliation{GSI Helmholtzzentrum f\"{u}r Schwerionenforschung GmbH, 64291 Darmstadt, Germany}

\author{K.~Gulyuz}
\affiliation{National Superconducting Cyclotron Laboratory, East Lansing, Michigan 48824, USA}
\affiliation{Department of Physics, Central Michigan University, Mount Pleasant, MI 48859, USA}

\author{C.~Izzo}
\altaffiliation[Current address: ]{Fermilab}
\affiliation{National Superconducting Cyclotron Laboratory, East Lansing, Michigan 48824, USA}
\affiliation{Department of Physics \& Astronomy, Michigan State University, East Lansing, Michigan 48824, USA}

\author{C.~Langer}
\affiliation{Department of Energy Technology, University of Applied Science Aachen, Campus J\"{u}lich, 52428 J\"{u}lich, Germany}

\author{F.~Montes}
\affiliation{National Superconducting Cyclotron Laboratory, East Lansing, Michigan 48824, USA}
\affiliation{Joint Institute for Nuclear Astrophysics -- Center for the Evolution of the Elements, Michigan State University, East Lansing, Michigan 48824, USA}

\author{W.-J~Ong}
\affiliation{Lawrence Livermore National Laboratory, Livermore, California 94550, USA}

\author{D.~Puentes}
\affiliation{National Superconducting Cyclotron Laboratory, East Lansing, Michigan 48824, USA}
\affiliation{Department of Physics \& Astronomy, Michigan State University, East Lansing, Michigan 48824, USA}

\author{M.~Redshaw}
\affiliation{Department of Physics, Central Michigan University, Mount Pleasant, MI 48859, USA}
\affiliation{National Superconducting Cyclotron Laboratory, East Lansing, Michigan 48824, USA}

\author{R.~Ringle}
\affiliation{National Superconducting Cyclotron Laboratory, East Lansing, Michigan 48824, USA}

\author{R.~Sandler}
\affiliation{National Superconducting Cyclotron Laboratory, East Lansing, Michigan 48824, USA}
\affiliation{Department of Physics, Central Michigan University, Mount Pleasant, MI 48859, USA}

\author{H.~Schatz}
\affiliation{National Superconducting Cyclotron Laboratory, East Lansing, Michigan 48824, USA}
\affiliation{Department of Physics \& Astronomy, Michigan State University, East Lansing, Michigan 48824, USA}
\affiliation{Joint Institute for Nuclear Astrophysics -- Center for the Evolution of the Elements, Michigan State University, East Lansing, Michigan 48824, USA}

\author{S.~Schwarz}
\affiliation{National Superconducting Cyclotron Laboratory, East Lansing, Michigan 48824, USA}

\author{C. S. Sumithrarachchi}
\affiliation{National Superconducting Cyclotron Laboratory, East Lansing, Michigan 48824, USA}

\author{A. A.~Valverde}
\altaffiliation[Current address: ]{Argonne National Laboratory, Lemont, IL 60439, USA}
\affiliation{Department of Physics, University of Notre Dame, Notre Dame, IN 46556, USA}

\author{I. T.~Yandow}
\affiliation{National Superconducting Cyclotron Laboratory, East Lansing, Michigan 48824, USA}
\affiliation{Department of Physics \& Astronomy, Michigan State University, East Lansing, Michigan 48824, USA}

\date{\today}

\begin{abstract} 
We performed a Penning trap mass measurement of $^{61}{\rm Zn}$ at the National Superconducting Cyclotron Laboratory and {\tt NuShellX} calculations of the $^{61}{\rm Zn}$ and $^{62}{\rm Ga}$ structure using the GXPF1A Hamiltonian to obtain improved estimates of the $^{61}{\rm Zn}(p,\gamma)^{62}{\rm Ga}$ and $^{60}{\rm Cu}(p,\gamma)^{61}{\rm Zn}$ reaction rates. Surveying astrophysical conditions for type-I X-ray bursts with the code {\tt MESA}, implementing our improved reaction rates, and taking into account updated nuclear masses for $^{61}{\rm V}$ and $^{61}{\rm Cr}$ from the recent literature, we refine the neutrino luminosity from the important mass number $A=61$ urca cooling source in accreted neutron star crusts. This improves our understanding of the thermal barrier between deep heating in the crust and the shallow depths where extra heat is needed to explain X-ray superbursts, as well as the expected signature of crust urca neutrino emission in light curves of cooling transients.

\end{abstract}

\maketitle

\section{Introduction and Motivation}

Accreting neutron stars provide unique probes of matter at high density and relatively modest temperature~\cite{Meis18}. These dense remnants of stellar explosions siphon matter into an accretion disk, while viscous interactions in the disk ultimately dump the (often) hydrogen and helium-rich fuel onto the neutron star surface. Fuel build-up drives the neutron star outer layers, consisting of a gaseous atmosphere on top of a liquid ocean that lays above a crystalline crust, out of thermal equilibrium and results in a number of astronomical observables. The frequently recurring explosions powered by hydrogen and helium-burning in the atmosphere are known as type-I X-ray bursts, while the much more powerful, and far less frequent, explosions originating from carbon fusion in the ocean are known as X-ray superbursts~\cite{Gall21}. Another class of X-ray transients, known as cooling transients, result when a prolonged episode of accretion ceases for some time and the crust relaxes to thermal equilibrium~\cite{Wijn17}.

Comparisons between observations of these phenomena and corresponding astrophysics model calculations can shed light on the properties of neutron star matter, such as the impurity of the crust, presence of superfluid neutrons, and neutron star compactness~\cite{Zamf12,Page13,Merr16,Deib17,Meis19}. Model-observation comparisons have identified the need for an as-of-yet-unexplained heat source located roughly 100~m below the neutron star surface, commonly referred to as ``shallow heating"~\cite{Coop09,Deib15,Pari17,Keek17}. However, model results are sensitive to the predicted thermal and compositional structure of the accreted crust, which is set by the nuclear reactions on and below the neutron star surface. 

Particularly consequential nuclides are known as urca pairs, where abundance is cyclically transferred between two nuclides within a spherical shell due to electron captures (EC) and $\beta^{-}$ decays. As a result, a substantial amount of energy is liberated from the crust by way of escaping neutrinos~\cite{Scha14}. Urca shells create thermal barriers, limiting heat transfer from deep layers of the crust into shallow layers where bursts are ignited, possibly leading to observable consequences~\cite{Deib16,Meis17}. Of these, urca cycling on nuclei with mass number $A=61$ produces the fourth largest neutrino luminosity $L_{\nu}$ of the roughly one-hundred $A$ present in X-ray burst ashes~\cite{Ong20}. Significant urca pairs include $^{61}{\rm Fe}$-$^{61}{\rm Mn}$, $^{61}{\rm Mn}$-$^{61}{\rm Cr}$, and $^{61}{\rm Cr}$-$^{61}{\rm V}$.

The $L_{\nu}$ from an urca pair is~\citep{Tsur70,Deib16}
\begin{equation}
L_{\nu} \approx L_{34}\times10^{34}{\rm{erg\,s}}{}^{-1}X(A)T_{9}^{5}\left(\frac{g_{14}}{2}\right)^{-1}R_{10}^{2} \ ,
\label{eqn:Lnu}
\end{equation}
where $X(A)$ is the mass fraction of the EC parent nucleus in the composition and $T_{9}$ is the temperature of the urca shell in units of $10^{9} \, \mathrm{K}$. $R_{10}\equiv R/(10~\rm{km})$, where $R$ is the radius of the urca shell from the neutron star center. $g_{14}\equiv g/(10^{14}~\rm{cm}\,\rm{s}^{-2})$, where $g = (GM/R^2)(1-2GM/Rc^2)^{-1/2}$ is the surface gravity of the neutron star with mass $M$, $c$ is the speed of light in vacuum, and $G$ is the gravitational constant.
$L_{34}(Z,A)$ is the intrinsic cooling strength of an urca pair with a parent nucleus that has $Z$ protons and $A$ nucleons:
\begin{equation}
L_{34}=0.87\left(\frac{10^{6}~{\rm{s}}}{ft}\right)\left(\frac{56}{A}\right)\left(\frac{Q_{\rm{EC}}}{4~{\rm{MeV}}}\right)^{5}\left(\frac{\langle F\rangle^{*}}{0.5}\right). \ 
\end{equation}
The energy-cost for EC is the EC $Q$-value $Q_{\rm EC}={\rm ME}(Z,A)-{\rm ME}(Z-1,A)$, where the atomic mass excesses ME are corrected by a Coulomb lattice energy $+C_{\ell}Z^{5/3}Q_{\rm EC,0}$, $Q_{\rm EC,0}$ is the $Q$-value without the lattice correction, and $C_{\ell}\approx3.407\times10^{-3}$~\cite{Roca08}. The factor $\langle F\rangle^{*}\equiv\langle F\rangle^{+}\langle
F\rangle^{-}/(\langle F\rangle^{+}+\langle F\rangle^{-})$, where the Coulomb factor $\langle F\rangle^{\pm}\approx2\pi\alpha
Z/|1-\exp(\mp2\pi\alpha Z)|$ and $\alpha\approx1/137$ is the
fine-structure constant. The comparative half-life of the weak transition $ft$ is the average for the $\beta$-decay and EC reactions in the urca cycle, $ft=(ft_{\beta}+ft_{\rm EC})/2$, where the two are related by the spin $J$ degeneracy of the initial states $ft_{\beta}/(2J_{\beta}+1)=ft_{\rm EC}/(2J_{\rm EC}+1)$~\cite{Paxt16}.

Neutron stars experiencing repeated X-ray bursts have crust compositions set by the burst ashes. While $X(A)$ can be modified by nuclear reactions in the crust, they largely reflect the surface values at depths relevant for urca cooling~\cite{Lau18}. The $^{61}{\rm Zn}(p,\gamma)^{62}{\rm Ga}$ reaction rate uncertainty has the most significant impact on $X(61)$ from X-ray burst model calculations when considering a range of astrophysical conditions, causing up to an order of magnitude change in the corresponding $L_{\nu}$, while $^{60}{\rm Cu}(p,\gamma)^{61}{\rm Zn}$ is also thought to be influential~\cite{Pari08,Cybu16}. Each of these reaction rates are presently poorly constrained. Commonly used theoretical estimates include the shell-model based rate of Ref.~\cite{Fisk01} and the Hauser-Feshbach rates from the codes {\tt NonSmoker}~\cite{Raus00} and {\tt Talys}~\cite{Koni08}. The former employed the code {\tt ANTOINE} with the KB3 interaction, known to have difficulty reproducing data above $A\approx52$~\cite{Honm04}, and presented unconverged results for levels above 4~MeV excitation energy. The latter two rely on statistical averages of nuclear properties and may not accurately describe a particular reaction rate of interest.

We present high-precision mass measurement results for $^{61}{\rm Zn}$ and improved $^{61}{\rm Zn}(p,\gamma)$ and $^{60}{\rm Cu}(p,\gamma)$ reaction rates based on these results and state-of-the-art shell-model calculations. Implementing this improved nuclear physics and other improvements involving neutron-rich $A=61$ nuclides from the recent literature for a range of astrophysical conditions in multi-zone X-ray burst model calculations, we refine estimates of $L_{\nu}$ from $A=61$ urca nuclei in accreted neutron star crusts. We describe our mass measurement in Section~\ref{sec:mass}, shell-model calculations and reaction rate results in Section~\ref{sec:rates}, and astrophysics model calculations in Section~\ref{sec:astro}. Results and implications for astrophysics are discussed in Section~\ref{sec:disc}, followed by concluding remarks in Section~\ref{sec:concl}. 

\section{Penning Trap Mass Measurements}
\label{sec:mass}

The mass measurement of $^{61}{\rm Zn}$ was performed with the Low Energy Beam and Ion Trap (LEBIT)~\cite{Ring13} facility at the National Superconducting Cyclotron Laboratory. A beam of $^{78}{\rm Kr}$ was accelerated to 150~MeV/nucleon by the Coupled Cyclotron Facility and impinged on a 672~mg/cm$^{2}$ beryllium target to produce a secondary beam via projectile fragmentation. $^{61}{\rm Zn}$ ions were selected using the A1900 fragment separator~\cite{Morr03} and deposited into a linear helium-filled gas-cell in the beam stopping facility~\cite{Sumi20}, which was operated at a pressure $\sim 100$~mbar. The ions were extracted through a radiofrequency quadrupole ion-guide, accelerated to 30~keV, and purified through a dipole magnet to provide mass-to-charge $(A/Q)$ separation. An insertable silicon detector was used to measure activity after the separator. Most activity was identified at $A/Q = 61$ corresponding to singly charged $^{61}{\rm Zn}$. The $^{61}{\rm Zn}^{+}$ ions were then sent to the LEBIT facility.

Within LEBIT, $^{61}{\rm Zn}^+$ ions were cooled and collected in the cooler buncher, a two-stage helium-gas filled Paul trap~\cite{Schw16}, before being released as short, low-emittance pulses towards LEBIT's 9.4~T Penning trap mass spectrometer~\cite{Ring09}. Before entering the trap, the pulses were further purified using a time-of-flight filter to only allow ions with $A/Q = 61$ to enter the trap. Ions were injected off-axis using a Lorentz steerer~\cite{Ring07} to prepare them with an initial magnetron orbit. During trapping, additional purification of isobaric contaminants was achieved using the targeted dipole radiofrequency (RF) excitation technique as well as the stored waveform inverse Fourier transform (SWIFT) method~\cite{Mars85,Blau04,Kwia15}. 

\begin{figure}[ht!]
\begin{center}
\includegraphics[width=1.0\columnwidth]{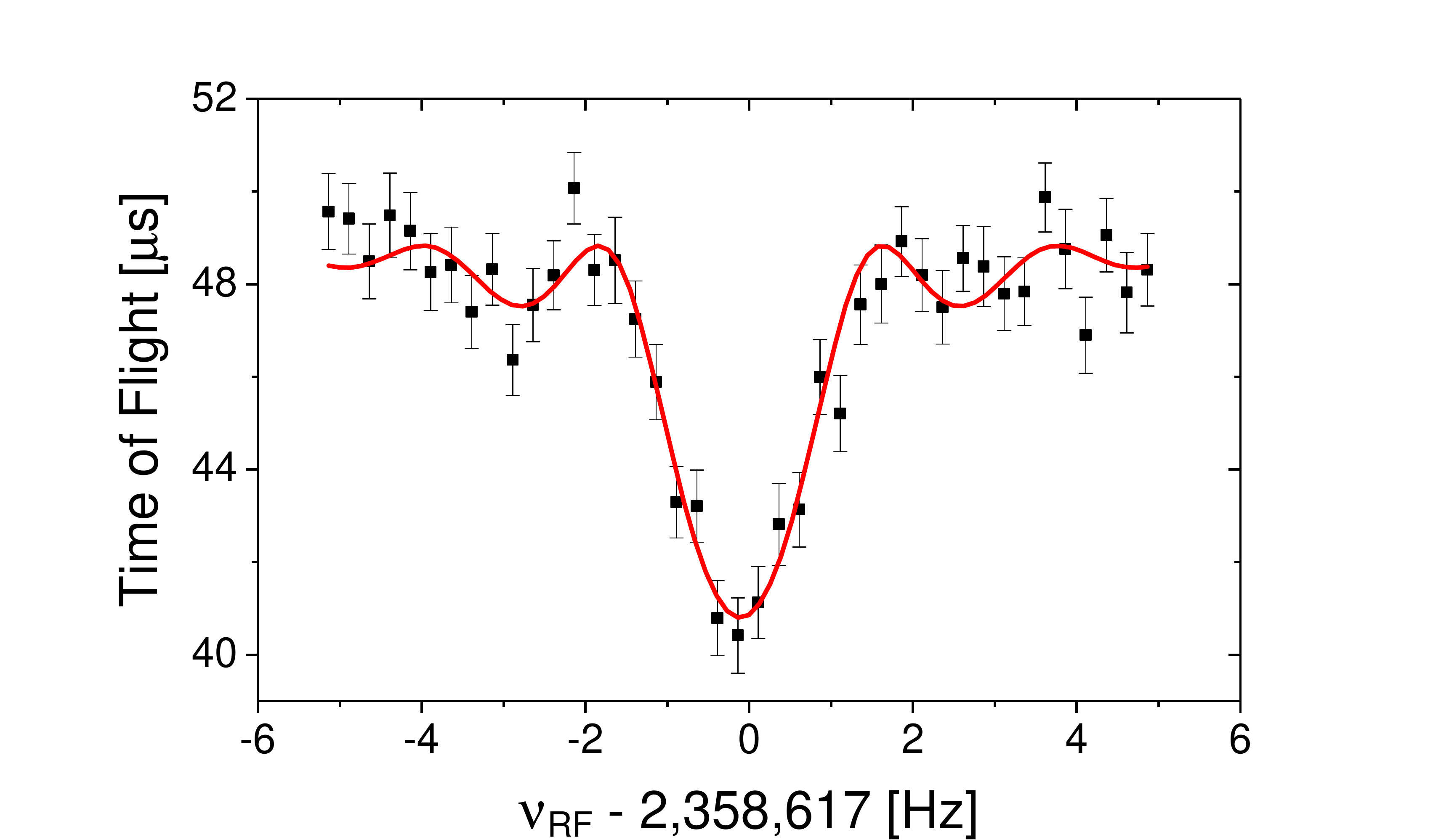}
\caption{ 
A $^{61}{\rm Zn}^+$ TOF-ICR resonance performed with a 500~ms excitation time. The red curve is a theoretical fit to the data~\cite{Koni95}. The minimum of the fit corresponds to the cyclotron frequency $\nu_c$.} \label{fig:TOF}
\end{center}
\end{figure}

The mass $m$ of an ion with charge $q$ was determined by measuring the ion's cyclotron frequency $\nu_{\rm c}=qB/2\pi m$ in a magnetic field of strength $B$. The cyclotron frequency was measured using the time-of-flight ion cyolotron resonance (TOF-ICR) technique~\cite{Graf80,Koni95,Dill18}. In this technique, a quadrupole RF excitation close to the cyclotron frequency, $\nu_{\rm RF} \approx \nu_{\rm c}$, is used to convert the initial magnetron motion into reduced cyclotron motion resulting in a large kinetic energy gain. The energy gain leads to a reduced time of flight through the magnetic field gradient after the ions are released from the Penning trap, which was measured with a microchannel plate (MCP) detector outside of the magnetic field. The cyclotron frequency was determined by scanning the excitation frequency $\nu_{\rm RF}$ and recording the TOF, resulting in a resonance response as shown in Figure~\ref{fig:TOF}. A theoretical fit to the data~\cite{Koni95} provides the minimum in the TOF response which corresponds to $\nu_{\rm c}$. The magnetic field strength was determined by taking TOF-ICR measurements of a reference ion produced in the gas cell, \textsuperscript{12}C\textsubscript{5}\textsuperscript{1}H\textsuperscript{+}, before and after each measurement of $^{61}{\rm Zn}^+$. The reference measurements were used to linearly interpolate the magnetic field strength to the time of the $^{61}{\rm Zn}^+$ measurement.

The experimental quantity of interest for Penning trap mass spectrometry is the frequency ratio, $R = \nu^{\rm int}_\text{ref}/\nu_{c}$, where $\nu^{\rm int}_\text{ref}$ is the interpolated cyclotron frequency from the \textsuperscript{12}C\textsubscript{5}\textsuperscript{1}H\textsuperscript{+} measurements. A series of four measurements of $R$ were taken over the course of two hours. $^{61}{\rm Zn}^+$ was measured with a 500~ms excitation time while \textsuperscript{12}C\textsubscript{5}\textsuperscript{1}H\textsuperscript{+} was measured with a 1~s excitation time. The resulting weighted average of these measurements was $\bar{R} = 0.998\, 880\, 050 (88)$ with a Birge ratio of 1.19(24)~\cite{Bir32}. Because the Birge ratio was greater than 1, the reported uncertainty in $\bar{R}$ has been inflated to account for any underestimated uncertainties.

Our mass measurement uncertainty is primarily statistical. The majority of the systematic uncertainties in $\bar{R}$ scale linearly with the difference in mass between the ion of interest and the reference. These systematic effects include spatial magnetic field inhomogeneities, electrostatic trapping field imperfections, and a possible misalignment between the trap and the magnetic field \cite{Bol90}. Because $^{61}{\rm Zn}^+$ and \textsuperscript{12}C\textsubscript{5}\textsuperscript{1}H\textsuperscript{+} are a mass doublet, these effects are negligible. Remaining systematic uncertainties affect the individual measurements of $R$. These include non-linear temporal fluctuations in the magnetic field, relativistic effects, and ion-ion interactions. The non-linear magnetic field fluctuations on $R$ are less than $1 \times 10^{-9}$ over an hour~\cite{Ring07A}. The total measurement time was two hours, making this uncertainty negligible. Relativistic effects were also negligible due to the large mass of the measured ions. Ion-ion interactions were  minimized by removing most of the contamination using SWIFT and dipole excitations and by limiting the analysis to include events with no more than five ions collected on the MCP, corresponding to eight or fewer ions in the trap based on the MCP efficiency of 63\%~\cite{Val15}. A count-rate class analysis was also performed to extrapolate the cyclotron frequency to the single ion level~\cite{Kel03}.

There was no evidence of $^{61}{\rm Zn}$ isomers in the TOF response during the measurements. The ion transport time to LEBIT and the excitation time within the Penning trap were sufficiently long enough that all isomeric contamination would be expected to decay prior to the TOF measurement. Therefore the measurement results presented here are for the $^{61}{\rm Zn}$ ground state.

From the ratio $\bar{R}$, we find ME($^{61}{\rm Zn}$)=-56355.5(5.0)~keV, which is a factor of three more precise than the most recent atomic mass evaluation (AME)~\cite{Huan21}. Electron and molecular binding energies have been ignored as their contributions to the mass are several orders of magnitude lower than the reported uncertainty. Using the AME results for the $^{60}{\rm Cu}$ and $^{62}{\rm Ga}$ masses, our new $Q$-values for $^{60}{\rm Cu}(p,\gamma)$ and $^{61}{\rm Zn}(p,\gamma)$ are 5299.2(5.3)~keV and 2920.5(5.1)~keV, respectively. These are to be compared to the AME $Q$-values of 5293(16)~keV and 2921(16)~keV, as well as the $Q$-values adopted by Ref.~\cite{Fisk01}: 5290.7~keV and 2278~keV.

\section{Reaction Rate Calculations}
\label{sec:rates}

In order to calculate astrophysical reaction rates, we estimated the level properties of $^{61}{\rm Zn}$ and $^{62}{\rm Ga}$ using the shell-model code {\tt NuShellX}~\cite{Brow14}. We employed the GXPF1A Hamiltonian~\cite{Honm04}, which was specifically tailored to describe the structure of $fp$-shell nuclei. For our calculations, the $fp$ model space has been truncated to allow up to one proton and one neutron hole in the $f_{7/2}$ orbital.

We calculated the $^{60}{\rm Cu}(p,\gamma)$ and $^{61}{\rm Zn}(p,\gamma)$ reaction rates using the narrow resonance approximation~\cite{Brun15}. Here, the astrophysical reaction rate is $N_{A}\langle\sigma v\rangle\propto\Sigma_{i}\omega\gamma_{i}\exp(-E_{r,i}/(k_{\rm B}T_{9}))$, where $k_{\rm B}$ is Boltzmann's constant and a resonance energy is $E_{r,i}=E^{*}_{i}-Q_{p,\gamma}$, calculated from the excitation energy $E^{*}_{i}$ of state $i$ and the $(p,\gamma)$ $Q$-value $Q_{p,\gamma}$. The resonance strength of state $i$ is 
\begin{equation*}
\omega\gamma_{i}=\frac{2J_{r,i}+1}{(2J_{p}+1)(2J_{\rm targ}+1)}\frac{\Gamma_{p,i}\Gamma_{\gamma,i}}{\left(\Gamma_{p,i}+\Gamma_{\gamma,i}\right)},
\end{equation*}
where $J_{\rm targ}$ is the spin of the proton-capture ``target" ($^{60}{\rm Cu}$ or $^{61}{\rm Zn}$) and the resonance spins $J_{r,i}$, partial proton-decay widths $\Gamma_{p,i}$, and partial $\gamma$-decay widths $\Gamma_{\gamma,i}$ are from the shell-model results.

\begin{figure}[ht!]
\begin{center}
\includegraphics[width=0.9\columnwidth]{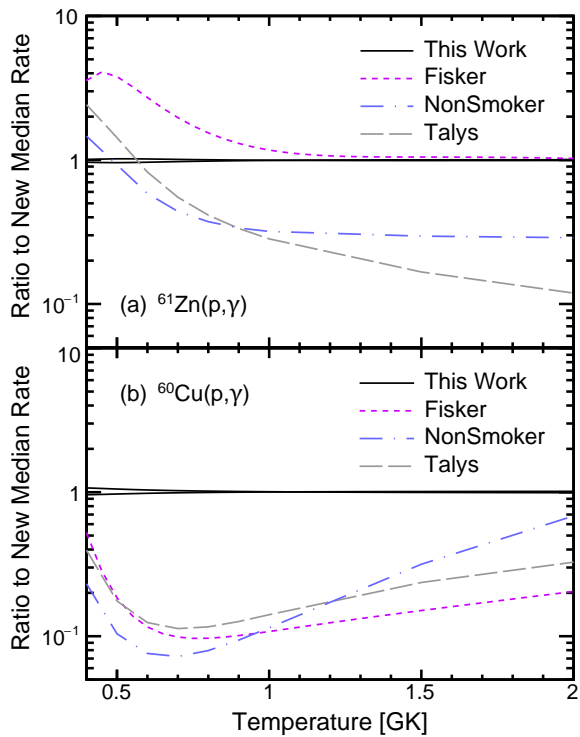}
\caption{ 
Ratio between the new (a) $^{61}{\rm Zn}(p,\gamma)$ and (b) $^{60}{\rm Cu}(p,\gamma)$ reaction rates and commonly adopted alternatives. {\tt Talys} rates were calculated using version 1.8 and a logarithmic energy binning for the continuum, with default settings otherwise. The median rate band reflects the uncertainty from the nuclear masses, including the impact on $\Gamma_{p}$.} \label{fig:RateRatio}
\end{center}
\end{figure}

\begingroup
\squeezetable
\begin{table}[t]
  \caption{\label{tab:rates}
New astrophysical reaction rates from this work. Rate units are cm$^{3}$mol$^{-1}$s$^{-1}$.}
 \def\arraystretch{1.25}
  \begin{ruledtabular}
  \begin{tabular}{ccc}
 Temperature [GK]	&	$^{60}{\rm Cu}(p,\gamma)$	&	$^{61}{\rm Zn}(p,\gamma)$	\\ \hline
0.10	&	7.01E-21	&	5.85E-22	\\
0.15	&	1.21E-16	&	1.58E-17	\\
0.20	&	6.77E-14	&	8.96E-15	\\
0.25	&	9.30E-12	&	1.78E-12	\\
0.30	&	5.32E-10	&	8.99E-11	\\
0.35	&	1.83E-08	&	1.81E-09	\\
0.40	&	3.70E-07	&	2.23E-08	\\
0.45	&	4.42E-06	&	2.09E-07	\\
0.50	&	3.42E-05	&	1.52E-06	\\
0.55	&	1.89E-04	&	8.66E-06	\\
0.60	&	8.02E-04	&	3.90E-05	\\
0.65	&	2.76E-03	&	1.43E-04	\\
0.70	&	8.04E-03	&	4.44E-04	\\
0.75	&	2.04E-02	&	1.20E-03	\\
0.80	&	4.64E-02	&	2.87E-03	\\
0.85	&	9.61E-02	&	6.24E-03	\\
0.90	&	1.84E-01	&	1.25E-02	\\
0.95	&	3.30E-01	&	2.34E-02	\\
1.00	&	5.59E-01	&	4.14E-02	\\
1.10	&	1.39E+00	&	1.12E-01	\\
1.20	&	2.99E+00	&	2.58E-01	\\
1.30	&	5.72E+00	&	5.30E-01	\\
1.40	&	9.97E+00	&	9.92E-01	\\
1.50	&	1.61E+01	&	1.72E+00	\\
1.60	&	2.46E+01	&	2.81E+00	\\
1.70	&	3.57E+01	&	4.35E+00	\\
1.80	&	4.96E+01	&	6.46E+00	\\
1.90	&	6.65E+01	&	9.24E+00	\\
2.00	&	8.65E+01	&	1.28E+01	\\
2.25	&	1.50E+02	&	2.60E+01	\\
2.50	&	2.33E+02	&	4.64E+01	\\
2.75	&	3.31E+02	&	7.53E+01	\\
3.00	&	4.41E+02	&	1.14E+02	\\
3.25	&	5.60E+02	&	1.62E+02	\\
3.50	&	6.85E+02	&	2.19E+02	\\
3.75	&	8.13E+02	&	2.86E+02	\\
4.00	&	9.41E+02	&	3.61E+02	\\
4.50	&	1.19E+03	&	5.31E+02	\\
5.00	&	1.43E+03	&	7.21E+02	\\
6.00	&	1.83E+03	&	1.12E+03	\\
7.00	&	2.13E+03	&	1.52E+03	\\
8.00	&	2.35E+03	&	1.87E+03	\\
9.00	&	2.49E+03	&	2.17E+03	\\
10.00	&	2.58E+03	&	2.41E+03	\\

  \end{tabular}
  \end{ruledtabular}
\end{table}
\endgroup

The resulting reaction rates (see Table~\ref{tab:rates}) are compared to rates commonly adopted in astrophysics model calculations in Figure~\ref{fig:RateRatio}. At the $\sim$0.5~GK temperatures relevant for urca nuclide production~\cite{Merz21}, our $^{60}{\rm Cu}(p,\gamma)$ reaction rate is nearly ten times larger than these other rate estimates. Our $^{61}{\rm Zn}(p,\gamma)$ rate generally falls between the Hauser-Feshbach rates and the shell-model rate of Ref.~\cite{Fisk01}. At $\sim$0.5~GK, our rate is roughly four times smaller than the Ref.~\cite{Fisk01} rate, which is currently recommended in the REACLIB database~\cite{Cybu10}, but is in closer agreement with Hauser-Feshbach rates.

\section{Astrophysics Model Calculations}
\label{sec:astro}

We assessed the impact of our newly determined reaction rates on $X(61)$ in X-ray burst ashes using the code {\tt MESA}~\cite{Paxt11,Paxt13,Paxt15} version 9793, following the calculations described in Ref.~\cite{Meis19}. These are multi-zone X-ray burst models that consist of an $\sim$0.01~km thick envelope that is heated from below by 0.1~MeV per accreted nucleon and treated as if it is atop a neutron star with mass $M=1.4~M_{\odot}$ and radius $R_{\rm NS}=11.2$~km ($\approx R$). We used nuclear reaction rates from REACLIB~\citep{Cybu10} version 2.2 within the 304~isotope network of Ref. \cite{Fisk08}, but adopted $^{60}{\rm Cu}(p,\gamma)$ and $^{61}{\rm Zn}(p,\gamma)$ rates from our work, Ref.~\cite{Fisk01}, or Ref.~\cite{Raus00}.

To survey the impact of our results over a range of astrophysical conditions, we performed calculations for three models, determining the ash composition after 18 bursts each. For Model A, found to reproduce observations of the year 2007 burst epoch of the source GS 1826-24~\cite{Meis18b}, the accretion rate is $\dot{M} = 0.17~\dot{M_{\rm E}}$ where $\dot{M_{\rm E}}=1.75\times10^{-8}~M_\odot$\,yr$^{-1}$ is the Eddington accretion rate from Ref.~\cite{Scha99}, with accreted hydrogen mass fraction $X_{\rm H}=0.70$, accreted metal mass fraction $Z=0.02$ with the solar metal distribution of Ref.~\cite{Grev98}, and accreted helium mass fraction $Y=1-X_{\rm H}-Z$. For Model B, which has a rapid proton-capture ($rp$)-process end-point near $^{80}{\rm Zr}$~\cite{Meis19}, the accretion rate is instead $0.05~\dot{M_{\rm E}}$. Model C is like Model A except $X_{\rm H}=0.75$, the upper limit in hydrogen-richness that one would expect due to Big Bang nucleosynthesis mass-fractions, in order to result in extended hydrogen burning.

\begin{figure}[ht!]
\begin{center}
\includegraphics[width=1\columnwidth]{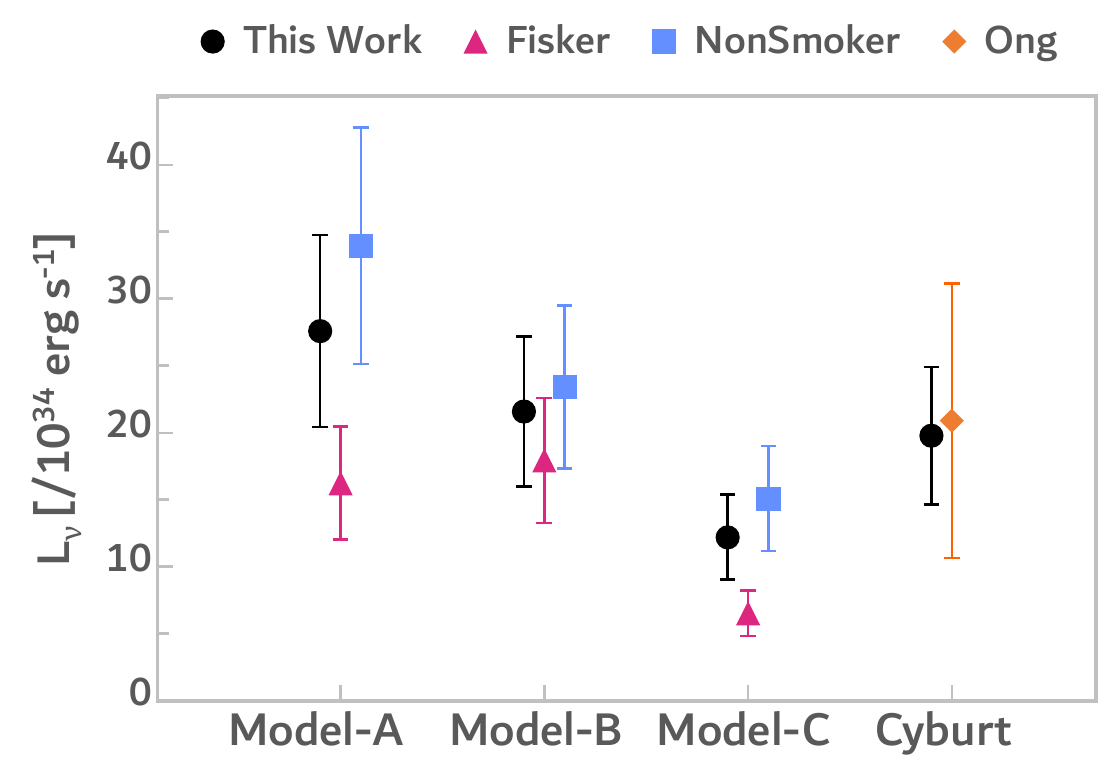}
\caption{ 
$L_{\nu}$ from $A=61$ for our results compared to the results of Ref.~\cite{Ong20} (Ong).  For the Cyburt column, $X(61)$ from Ref.~\cite{Cybu16} was used.} \label{fig:Lnu}
\end{center}
\end{figure}

We computed $L_{\nu}$ from the significant $A$=61 urca pairs using $X(61)$ from Models A--C, the rough upper temperature limit for an accreted crust at urca shell depths $T_{9}$=1~\cite{Meis17}, $R$=11.2~km, and $M$=1.4~$M_{\odot}$. We used ME from Ref.~\cite{Huan21} to calculate $Q_{\rm EC}$ for the $^{61}{\rm Fe}$-$^{61}{\rm Mn}$, $^{61}{\rm Mn}$-$^{61}{\rm Cr}$, and $^{61}{\rm Cr}$-$^{61}{\rm V}$ urca pairs and to update $ft_{\beta}$ from Refs.~\cite{Radu13,Craw09,Ong20} using the {\tt LOGFT} tool from the National Nuclear Data Center. Relative to Ref.~\cite{Ong20}, ME have been updated for $^{61}{\rm V}$~\cite{Meis20,Mich20} and $^{61}{\rm Cr}$~\cite{Moug18}. We calculated $ft$ from $ft_{\beta}$ using $J$ from Ref.~\cite{Kond21}. Figure~\ref{fig:Lnu} compares the sum of $L_{\nu}$ from these three urca pairs to the $L_{\nu}$ one obtains when using the nuclear physics data and $X(61)$ employed in Ref.~\cite{Ong20}. Uncertainty contributions to $L_{\nu}$ include error propagation of $\delta Q_{\rm EC}$, $\delta ft$, and $\delta X(61)$, where we take the latter to be 10\% based on an analysis of the burst-to-burst variability after the ash abundance distribution has converged. We emphasize that Ref.~\cite{Ong20} used $X(61)$ from Ref.~\cite{Cybu16}, which were calculated for a single-set of astrophysical conditions (similar but not identical to Model A) with the code {\tt KEPLER}~\cite{Weav78,Woos04}. Tentative comparisons between results from X-ray burst calculations performed with {\tt MESA} and {\tt KEPLER} show general agreement~\cite{Merz21,John20}, but rigorous benchmarking has not been performed to date.

\section{Discussion}
\label{sec:disc}

Figure~\ref{fig:Lnu} shows the impact on $L_{\nu}$ of our updated nuclear physics and X-ray burst calculations. The first three columns show the impact on $L_{\nu}$ of varied astrophysical conditions (Models A, B, and C) and of different estimates for the $^{60}{\rm Cu}(p,\gamma)$ and $^{61}{\rm Zn}(p,\gamma)$ reaction rates. The fourth column compares our results when using $X(61)$ calculated by Ref.~\cite{Cybu16} in order to isolate the impact of updating the nuclear physics details for neutron-rich $A=61$ nuclei.

We find that $L_{\nu}$ is generally similar to the results of Ref.~\cite{Ong20}, but can be roughly two times lower or one-third higher, depending on the X-ray burst model conditions. Hydrogen-rich conditions reduce $X(61)$. This can be attributed to prolonged hydrogen burning, which will enable $rp$-process nucleosynthesis of heavier nuclides. As our newly calculated $^{61}{\rm Zn}(p,\gamma)$ reaction rate falls between the predictions of Ref.~\cite{Fisk01} and Ref.~\cite{Raus00}, so too does $X(61)$. As expected, the higher $^{61}{\rm Zn}(p,\gamma)$ reaction rate of Ref.~\cite{Fisk01} more efficiently destroys $X(61)$, resulting in the lowest $L_{\nu}$ for a given set of astrophysical conditions. For the relatively hydrogen-deficient conditions of Model B, variations in the $^{61}{\rm Zn}(p,\gamma)$ reaction rate have less of an impact on $X(61)$. This is likely due to the absence of hydrogen to burn at late times in the burst when cooler burning would result in more of the $rp$-process flow passing through $^{61}{\rm Zn}$.

In test calculations where we only varied the $^{60}{\rm Cu}(p,\gamma)$ reaction rate, we found no impact on $X(61)$. This is contrary to the results of Ref.~\cite{Pari08} and the single-zone X-ray burst calculations of Ref.~\cite{Cybu16}. A likely explanation is that both the post-processing results of Ref.~\cite{Pari08} where $^{60}{\rm Cu}(p,\gamma)$ was found to have an impact and the single-zone model calculations of Ref.~\cite{Cybu16}  are designed to mimic burning in the hottest regions of a multi-zone X-ray burst (where most of the X-ray flux originates), whereas high-$A$ urca nuclide production happens at shallower depths~\cite{Merz21}. Focusing on only the hottest zone effectively results in a different flow of the $rp$-process and therefore a different reaction rate sensitivity. We note that the multi-zone X-ray burst calculation results of Ref.~\cite{Cybu16} also demonstrated a negligible impact of $^{60}{\rm Cu}(p,\gamma)$ rate variations on $X(61)$.

Updating the nuclear physics of $A=61$ nuclei in the neutron star crust slightly reduced $L_{\nu}$, while reducing the uncertainty by a factor of two. Figure~\ref{fig:Lnu} shows that the nuclear physics uncertainty contributions from the crust $A=61$ nuclei are now smaller than the uncertainty contribution to $L_{\nu}$ from unknown X-ray bursting conditions. The largest remaining nuclear physics uncertainty contributions for $A=61$ crust nuclei come from $ft$ for the $^{61}{\rm Mn}$ and $^{61}{\rm Fe}$ EC reactions, where each of these uncertainties are primarily due to the uncertainty in the ground-state branching of the corresponding $\beta$-decay. Additional uncertainty to $L_{\nu}$ may be contributed from the unknown $^{61}{\rm Ga}(p,\gamma)$ reaction rate, as well as uncertainty in the $^{61}{\rm Ga}$ mass~\cite{Meis19,Scha17}. Furthermore, experimental data on the $^{61}{\rm Zn}(p,\gamma)$ resonances are needed to improve constraints for this reaction rate.

The overall importance of $A=61$ urca cooling from X-ray burst ashes may be impacted by neutron-transfer reactions occurring in the crust~\cite{Chug19}. The first nuclear reaction network calculations exploring the impact of neutron transfer on urca cooling have shown that cooling can be reduced or enhanced, depending on the details of the $X(A)$ redistribution~\cite{Scha21}. However, the neutron transfer rates are still first-order estimates and sensitivities to nuclear and astrophysics uncertainties have yet to be explored. As such, no firm conclusions can yet be drawn on how neutron transfer reactions would impact the $L_{\nu}$ reported in this work.

Moreover, our results demonstrate that determining $L_{\nu}$ for a given accreting neutron star source depends sensitively on the surface burning history. Variations of the accretion rate over time will ultimately create a varying $X(61)$ with depth. As accretion continues, surface burning ashes will be buried and $X(61)$ within the urca shells, where the electron-Fermi energy is near $Q_{\rm EC}$ of the urca pairs, will vary over time. This demonstrates the need for consistent modeling of the surface-to-core phenomena of accreting neutron star sources.

\section{Conclusions}
\label{sec:concl}

In summary, we improved the nuclear physics constraints for $A=61$ nuclides involved in urca cooling in the crust of accreting neutron stars. Our Penning trap mass measurement of $^{61}{\rm Zn}$ and {\tt NuShellX} shell-model calculations of the $^{61}{\rm Zn}$ and $^{62}{\rm Ga}$ structure result in higher-fidelity estimates of the $^{60}{\rm Cu}(p,\gamma)$ and $^{61}{\rm Zn}(p,\gamma)$ reaction rates. The former rate is around ten times larger at temperatures relevant for urca nuclide production, while the latter is in between prior theoretical estimates. We implemented these reaction rates in {\tt MESA} X-ray burst models to calculate $X(61)$, used the recent literature to update nuclear properties of neutron-rich $A=61$ nuclides, and calculated the corresponding $L_{\nu}$ from urca cooling in a hot accreted crust. Our improved estimates for $L_{\nu}$ will refine calculations of observable phenomena from accreting neutron stars, in particular X-ray superbursts and cooling transients. The updated $^{60}{\rm Cu}(p,\gamma)$ and $^{61}{\rm Zn}(p,\gamma)$ reaction rates presented in this work may also impact calculation results for $\nu p$-process nucleosynthesis~\cite{Nish19}.

\begin{acknowledgments}
This work was funded by the U.S. Department of Energy through Grants No. DE-FG02-88ER40387, DE-NA0003909, DE-SC0019042, and DE-SC0015927. This work benefited from support by the U.S. National Science Foundation through Grant Nos. PHY-1430152 (Joint Institute for Nuclear Astrophysics -- Center for the Evolution of the Elements), PHY-1565546 (Operation of the NSCL as a National User Facility \& Research Program), PHY-1713857 (Nuclear Structure and Nuclear Astrophysics at the University of Notre Dame), and PHY-2111185 (The Precision Frontier at FRIB: Masses, Radii, Moments, and Fundamental Interactions).  
\end{acknowledgments}

\bibliographystyle{apsrev4-2}
\bibliography{References}

\end{document}